# Thin Flexible Lithium Ion Battery Featuring Graphite Paper Based Current Collectors with Enhanced Conductivity

Hang Qu [1], Jingshan Hou [1], Yufeng Tang [2], Oleg Semenikihin [3], and Maksim Skorobogatiy [1,*]

[1] Department of Physics Engineering, Ecole Polytechnique de Montreal, Montreal, Quebec, H3C 3A7, Canada.

[2] CAS Key Laboratory of Materials for Energy Conversion, Shanghai Institute of Ceramics, Chinese Academy of Sciences, Shanghai 200050, PR China.

[3] Department of Chemistry, Western University, London, Ontario, N6A 5B7, Canada.

**Abstract:** A flexible, light weight and high conductivity current collector is the key element that enables fabrication of high performance flexible lithium ion battery. Here we report a thin, light weight and flexible lithium ion battery that uses graphite paper enhanced with a nano-sized metallic layers as the current collector, $LiFePO_4$ and $Li_4Ti_5O_{12}$ as the cathode and anode materials, and PE membrane soaked in $LiPF_6$ as a separator. Using thin and flexible graphite paper as a substrate for the current collector instead of a rigid and heavy metal foil enables us to demonstrate a very thin Lithium-Ion Battery into ultra-thin (total thickness including encapsulation layers of less than 250 μm) that is also light weight and highly flexible.

## 1 Introduction

Many wearable and portable electronic devices require efficient, compliant power sources that can fully function when bent, folded, or compressed. [1] Lithium-ion batteries (LIBs) dominate the portable power-source market due to their high energy density, high output voltage, long-term stability and environmentally friendly operation. [1, 2, 3] High performance flexible LIBs are considered to be one of the most promising candidates of the power sources for the next generation flexible electronic devices. LIBs are typically consisting of several functional layers (**see Fig. 1(a)**). When battery flexibility is desired, all of the battery components should be flexible [1]. Among the various functional layers, the current collectors affect critically the battery performance, and their flexibility is typically difficult to achieve together with high conductivity.

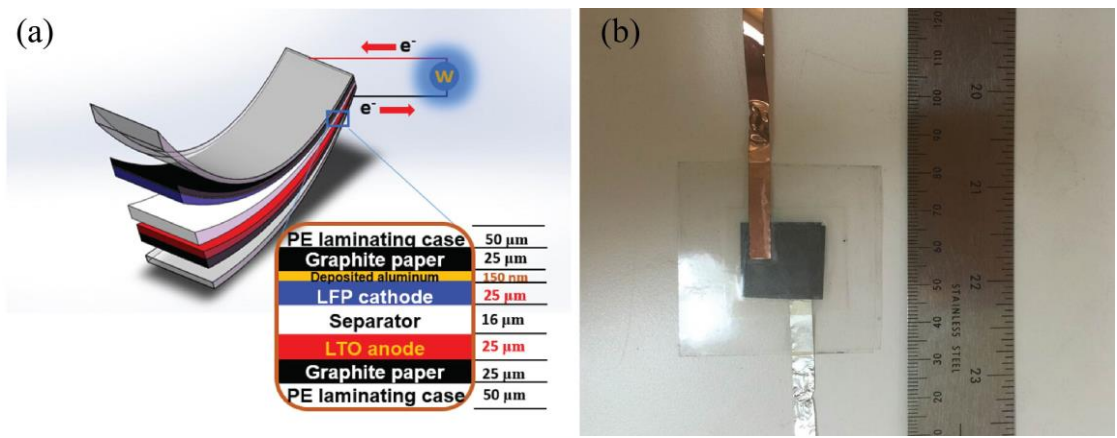

Fig.1 (a) Schematic of the thin, flexible lithium ion battery and (b) a lithium-ion battery sample

Many approaches have been explored for designing flexible LIBs.[4, 5, 6, 7, 8] Traditionally electrode active materials are coated onto a Cu foil which then works as anode, and an Al foil is generally used as cathode. The as-fabricated LIBs are typically heavy and rigid, which is not suitable for truly wearable applications. Recently, tremendous effort has been dedicated to the R&D of flexible current collectors and free standing electrodes based on carbon nanotube (CNT) or grapheme composites. Note that synthesis of these composites requires sophisticated processes, and cost of such materials is high (~1000$ per gram). Rigidity, heavy weight, price, etc. create significant barriers that prevent LIBs utilization in the wearable market. Therefore, highly conductive, light weight and flexible current collectors suitable for mass production are in high demand for the fabrication of light weight, flexible LIBs. In what follows, we report a flexible lithium ion battery using graphite-paper (GP) with enhanced conductivity as current collectors (**Fig. 1b**). The enhancement of conductivity of GP was achieved by depositing a sub-micron thick metal layer onto a commercial graphite paper by physical vapor deposition (PVD). Particularly, we use an aluminum-deposited GP as a current collector for cathode and a bare or a copper-deposited GP as a current collector for anode. Furthermore, LiFePO$_4$ (LFP) cathode and Li$_4$Ti$_5$O$_{12}$ (LTO) anode active materials are used in a PVDF flexible matrix together with a PE nanostructured membrane as a separator in order to build a flexible, high performance LIB.

## 2 Experimental

### 2.1 Chemicals and materials

LiFePO$_4$ (LFO), Li$_4$Ti$_5$O$_{12}$ (LTO) and Polyethylene (PE) separator were obtained from Targray Technology International Inc; Graphite paper (>99%; thickness: ~25μm) was purchased from Suzhou Dasen Electronic; Multi-walled carbon nanotube (MWCNT) and copper powder (> 99.7%) were obtained from Sigma-Aldrich. Al pellets (99.99%) are obtained from Kurt J. Lesker.

**2.2 Battery sample fabrication**

Highly conductive and flexible graphite-based current collector was fabricated by depositing a sub-micron thick aluminum or copper layer onto a graphite sheet by PVD (evaporative deposition). The obtained conductivity-enhanced GP (abbreviated as Al@GP and Cu@GP in the following) current collector was then used as substrates for deposition of LFP/PVDF and LTO/PVDF to make LFP-Al@GP and LTO-Cu@GP electrodes.

To synthesize anode and cathode, LFP and LTO were firstly pre-mixed with MWCNT in a mortar, respectively. Then mixture was then dispersed in Polyvinylidene fluoride (PVDF)/1-Methyl-2-pyrrolidone solution by a magnetic stirrer for 4 hours. The optimal weight ratio of LFP (or LTO), MWCNT and Polyvinylidene fluoride (PVDF) is found to be 8:1:1. The obtained slurry was then poured onto the current collector and made into a uniform wet-film using a Micrometer Adjustable Film Applicator (MTI Cooperation). The composites were dried in a vacuum furnace at 80 ℃ for 3h, and were then cut into 2*2 $cm^2$ square shape electrodes. These electrodes are then further dried at 110 ℃ for 12 h to ensure that the solvent (1-Methyl-2-pyrrolidone) in slurry is completely evaporated.

Batteries were finally assembled by stacking the as-prepared electrodes and the PE separator layer. 3 droplets (~0.15 mL) of electrolyte which is 1.0 M LiPF$_6$ dissolved in ethylene carbonate/dimethyl carbonate (EC/DEC) = 50/50 (v/v) from Sigma-Aldrich were added to the battery during the stacking process. Note that the battery assembly process was completed in a N$_2$-filled glove box to avoid oxidation of electrolyte. Finally, the battery is encapsulated with PE films using a lamination machine before it is taken out for characterization.

## 2.3 Characterization

The image of the cross section of the battery sample was taken by an optical microscope. The charge/discharge tests and EIS measurements were performed by an IviumStat.XR Electrochemcial Work Station. All of the capacities and C-rate currents in this work were calculated based on the mass of LFP active materials (1 C corresponding to 170 mAh/g).

## 3 Results and Discussion

### 3.1 Electrochemical performance of the LFP/Al@GP electrode.

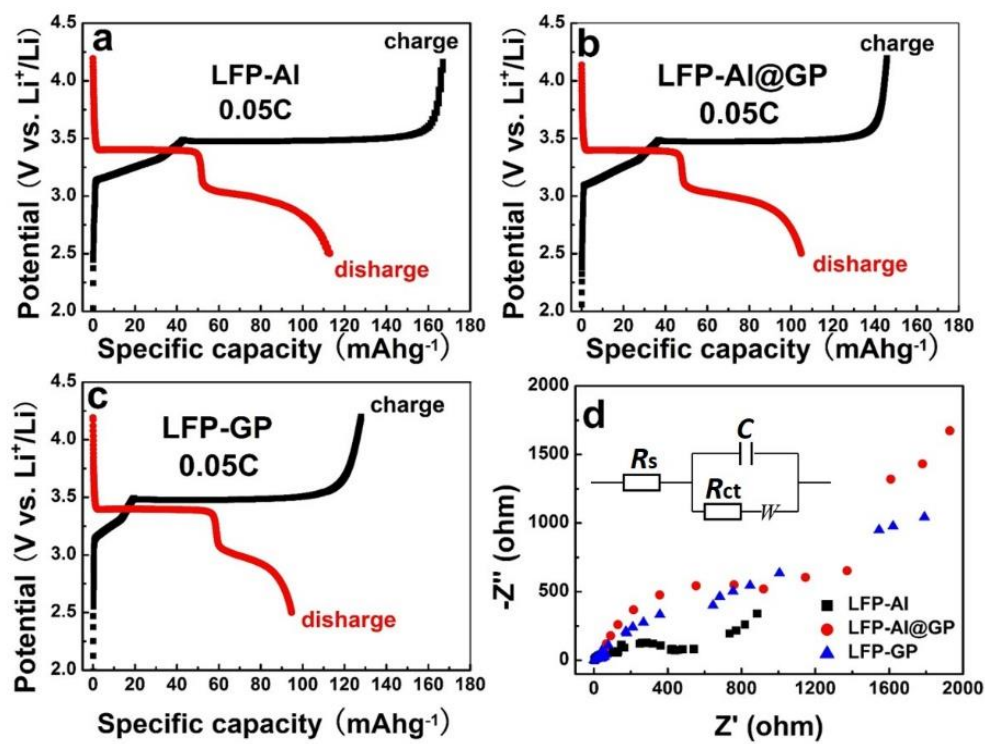

**Fig. 2** charge/discharge voltage curves of the half cells with rate of 0.05 C. a) LFP-Al (foil), b) LFP-Al@GP (Al enhanced GP), and c) LFP-GP (pure GP). d) Equivalent circuit and electrochemical impedance spectra of the LFP-Al@GP electrode, LFP-Al@GP electrode and a reference LFP-Al current collector.

The electrochemical performance of the LFP-Al@GP electrode was characterized by the galvanostatic charge-discharge tests and EIS measurements. A half-cell battery with an active area of 2*2 cm$^2$ was fabricated using LFP-Al@GP as the cathode and pure GP directly as the anode. The performance of this half-cell battery was compared with its two counterparts in which the cathode is fabricated by coating LFP onto an Al foil current collector (abbreviated as LFP-Al) or by coating LFP onto a GP current collector (abbreviated as LFP-GP). Note that the anode of reference half-cell batteries is also a 2*2 cm$^2$ sized GP, the same as used in the LFP-Al@GP battery. The lithium-ion insertion/extraction properties of the LFP-Al cathode and LFP-Al@GP cathode were investigated by galvanostatic charge-discharge measurements. **Fig. 2** represents the voltage curves of the first cycle of the charge/discharge test for batteries using the LFP-Al (**Fig. 2a**), LFP-Al@GP (**Fig. 2b**) and LFP-GP (**Fig. 2c**) with a 0.05 C charge/discharge rate. All the half-cells feature a ~ 3.2 V open circuit voltage. The capacity of the LFP-Al@GP//GP cell at 0.05 C discharge rate is 104.8 mAh/g with a coulombic efficiency of 71 % (see **Fig.2a**). This is comparable to the performance of the half-cell battery using LFP-Al cathode that has a capacity of 113 mAh/g and a coulombic efficiency of 67.7 % (see **Fig 2b**). These results indicated that the Al@GP could be an excellent current collector for the cathode as its electrochemical properties are virtually identical to pure Al foil electrodes.

To investigate the conductivity of a cathode, EIS measurement of LFP-Al@GP was performed, and the results were compared to those of LFP-Al and LFP-GP. **Fig. 2d** shows the Nyquist plots of the fresh LFP-Al//GP, LFP-Al@GP//GP, and LFP-GP//GP half-cells after the first 0.2C cycle. Nyquist plots are composed of a depressed semicircle in the high-to-medium frequency region together with a slope in the low frequency region. According to the order of decreasing frequency, the EIS spectra can be divided into three distinct regions. The first intercept on the real axis in **Fig.2d** (high frequency region) gives the equivalent series resistance (ESR), $R_s$,, which is a bulk electrolyte resistance. The second intercept (lower frequency region) gives a sum of the electrolyte resistance, Rs, and the charge transfer resistance, $R_{ct}$, which is the electrode/electrolyte interfacial resistance.[15]. The values of the charge-transfer resistance $R_{ct}$ of the LFP-Al@GP electrode was 920 Ω, which are higher than that of the LFP-Al ($R_{ct}$ = 400 Ω), but is significantly lower than that of the LFP-GP. Note that the Al@GP is much more flexible and lighter compared to pure Al foil. All these results indicate that the Al@GP fabricated by PVD method, could be an excellent candidate of current collector for highly flexible LIBs. However, it is meaningless to discuss more of the batteries' performance when GP was used directly as the anode without any modification or improvements.

It should be mentioned that, if GP is directly used as a current collector, coating of the battery active materials directly onto GP is challenging, since the coated layer peels off easily. Thus, yield of the corresponding electrodes is less than 10 %. However, yield of the electrodes can experimentally reach almost 100%, if electrodes are made from the GP enhanced with a nano-sized metal layers. Therefore, depositing a submicron thick metallic layer onto GP not only enhances the conductivity of the current collector, but also mechanically improves the adhesion between the battery active materials and the current collector, thus facilitating the subsequent battery assembly process and its reliability.

**3.2 Assembly and Electrochemical Behavior of the LFP-Al@GP//LTO-GP Flexible full Battery**

Since LFP and LTO have the well-matched theoretical capacities of 170 mAh/g and 175 mAh/g，respectively, and the LFP/Al@GP electrode showed excellent electrochemical performance that is comparable to that of LFP-Al, we proceeded with a full battery assembly. We fabricated a thin, and flexible LFP-Al@GP//LTO-GP full battery using the flexible LFP-Al@GP cathode and the LTO-GP anode.

To assemble the full battery, the integrated cathode (or anode electrode) was first attached to the sticky side of a polyethylene (PE) lamination film. Then we stacked the PE separator on top of the cathode, and added three droplets of $LiPF_6$ electrolyte. The counter electrode and another PE lamination film were then stacked to the battery structure one after another. Finally, the full battery was then encapsulated by a laminating machine. Note that all the assembly process was carried out in a $N_2$-filled glove box.

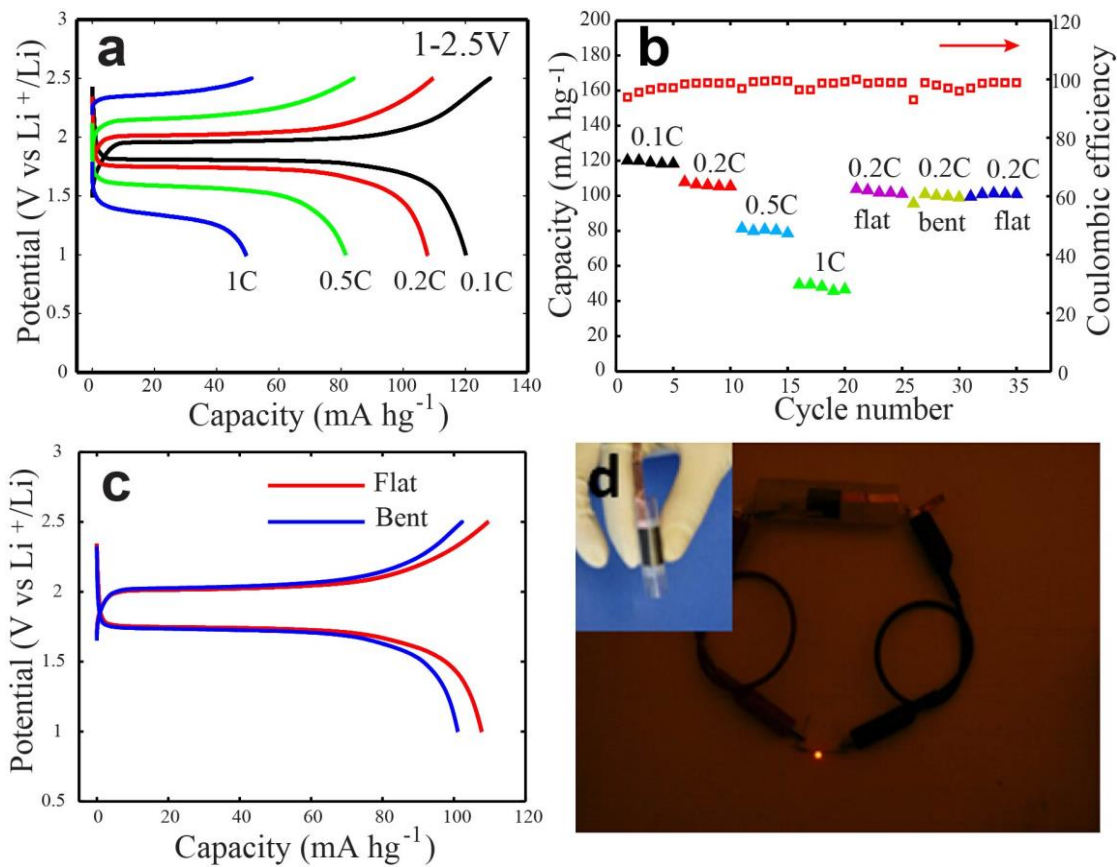

**Fig. 3** Characterization of the full flexible LFP-Al@GP//LTO-GP full battery. (a) Charge/discharge voltage curves of the battery with different rates. (b) Capacity and coulombic efficiency of the battery with different rates for 35 cycles. (c) Galvanostatic charge/discharge curves of the flat battery (red line) and a bent battery (blue line). The bending radius of the battery is ~1 cm. (d) A LED is lit up by a bent full battery. The inset demonstrates the excellent flexibility of the full battery.

The electrochemical properties of a full battery were investigated using charge /discharge cyclic analysis with different charge/discharge rates (from 0.1 C to 1 C) as shown in **Fig. 3a.** Consistent with the theoretical value of a LFP-LTO battery, the operating voltage of this full battery was found to be ~1.9 V. As shown in **Fig. 3a** and **b**, the LFP-Al@GP//LTO-GP battery has an excellent rate capacity, which is calculated to be 120.2, 107.8, 81.8 and 49.4 mAh g$^{-1}$ at 0.1, 0.2, 0.5, and 1 C, respectively. It can be clearly seen that the discharge capacities exhibit a tendency to decrease with the increment of current density; however, the voltage plateau still remains flat even when current density is up to 1C, indicating an excellent charge/discharge rate performance. Moreover, after a series of tests under different charge/discharge rates, the discharge capacity of the battery is still as high as 103.8 mAhg$^{-1}$, when the rate turned back to 0.2 C. This suggests that the structure of each components of the full battery remains intact after subjecting to high current densities. The coulombic efficiencies of the full battery were > 94 % during the whole cycle.

Due to the small thickness and high flexibility of the metal-enhanced GP current collectors and the as-integrated LFP-Al@GP and LTO-GP electrodes, the full battery shows excellent flexibility. The effect of bending on the performance of the flexible battery was also investigated. **Fig. 3c** shows the charge/discharge performance of the bent battery at 0.2 C after 25 charge/discharge cycles under different rates. Compared with the flat state, only a slight overpotential was observed, and a 2 % of capacity decrease was found. What's more, the flexible battery shows an excellent cyclic stability both under flat and bent states (see **Fig. 3b and c)**. The retention capacity of the bent battery is ~92% of that of the flat battery even after the battery undergoes 25 charge/discharge cycles. Furthermore, we note that even after bending the battery for another 5 charge/discharge cycles, more than 93% of the capacity can be still restored if the battery is released. In **Fig. 3d**, we demonstrate that a LED could be lit up by the battery bent at a 1 cm radius.

## 3.3 Assembly and Electrochemical Behavior of the LFP-Al@GP//LTO-Cu@GP Flexible Full Battery

LFP-Al@GP//LTO-Cu@GP full battery was also assembled by using the flexible LFP-Al@GP cathode and the LTO-Cu@GP anode. The assembly process is the same in the case of the above-mentioned LFP-Al@GP//LTO-GP battery, except we now replace the LTO-GP anode by LTO-Cu@GP. Therefore, a sub-micron thick Cu layer was deposited onto GP using physical vapor deposition method.

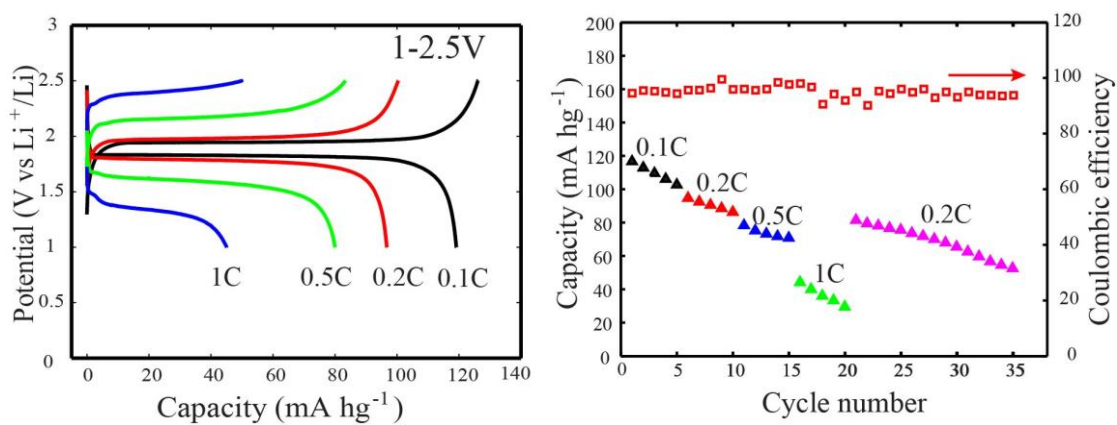

Fig. 4. Charge/discharge measurements of an LFP-Al@GP// LTO-Cu@GP battery. (a) Charge/discharge voltage curve of the battery as a function of the battery capacity. (b) Capacity and coulombic efficiency of the battery with different charge/discharge rates for 35 cycles.

Charge/discharge measurements are also performed for this full battery. The working voltage of the battery is also ~ 1.9 V (**Fig. 4**). The rate capacity of this battery is measured to be ~ 116.1, 96.4, 79.9 and 44.9 mAhg$^{-1}$, respectively. The capacities of this full battery are comparable to those of the LFP-Al@GP// LTO-GP battery. After 20 cycles of charge/discharge tests under different rates, the capacity of the battery could still achieve 81.5 mAhg$^{-1}$. We note that the capacity of this battery degrades faster during the cyclic charge/discharge tests, as compared to that of the LFP-Al@GP// LTO-GP battery. This is may be due to the oxidation of the electrolyte, which and future investigation is in order. The coulombic efficiency of the battery in all of the charge/discharge tests is above 91%.

**4 Conclusions**

In summary, an ultra-thin flexible battery with total thickness of less than 250 μm was successfully fabricated by using conductivity-enhanced metal-deposited GP current collector. To fabricate this highly flexible and conductive current collector, we deposit sub-micron thick metallic layers onto the GP using PVD technique. Compared to traditional current collectors based on pure metal (such as Al or copper) foils, the proposed current collector is advantageous due to its light weight, high flexibility and improved surface adhesion for depositing battery electrode materials. The battery uses $LiFePO_4$ and $Li_4Ti_5O_{12}$ as the cathode and anode materials, and PE membrane soaked in $LiPF_6$ as a separator. The battery could achieve a rate capacity of ~100 mAhg$^{-1}$ under standard 0.2C charge/discharge rate. Besides, the battery could retain its capacity even after intensive cyclic charge/discharge operation. During all the battery operation, the coulombic efficiency of the battery remains above 91%. We believe that this battery could find its niche markets in numerous fields relevant to portable or wearable electronic devices.